%% file: main.tex
\documentclass{llncs}
\usepackage{makeidx}  % allows for indexgeneration
\usepackage{graphicx}
\usepackage{cite}
\begin{document}
\title{Bringing Together Project Simulation and Psychometric Tests: A Business Game Proposal}
\titlerunning{A Business Game Proposal}  % abbreviated title (for running head)
\author{
Micha{\l} Okulewicz\inst{1},
Weronika Aniper\inst{2},
Bart{\l}omiej Dach\inst{1},
Piotr Filarski\inst{1},
Piotr Jenczyk\inst{4},
Julita O{\l}tusek\inst{3}}
\authorrunning{Micha{\l} Okulewicz et al.} % abbreviated author list (for running head)
\institute{Warsaw University of Technology,\\Faculty of Mathematics and Information Science,\\
\email{M.Okulewicz@mini.pw.edu.pl}
\and
Warsaw University of Technology,\\Faculty of Management
\and
Warsaw University of Technology,\\Faculty of Electronics and Information Technology
\and
Warsaw University of Technology,\\Faculty of Mechatronics}

\maketitle              % typeset the title of the contribution

\begin{abstract}
%The abstract should summarize the contents of the paper
%using at least 70 and at most 150 words. It will be set in 9-point
%font size and be inset 1.0 cm from the right and left margins.
%There will be two blank lines before and after the Abstract. \dots
This article identifies a gap between the existence of a various psychometric tests approaches and other team performance assessment tools (e.g. business and management games). As a response to the lack of tools able to utilize the knowledge about the behaviour patterns expressed in work environment and descriptive business games, the article presents a game proposal. The mechanics of the proposed game should allow for a quantifiable assessment of an individual in the simulated project development environment, making it possible to evaluate not only the team's performance, but the impact of actions of the individual players and their interactions.

\end{abstract}

\section{Introduction}

\input{01_introduction/introduction.tex}

\section{Related Work}
\label{sec:related-work}
The following section presents the applicability of business games  and usage of the psychometric tests in assessment of the project developing teams.

\subsection{Business Games}
\label{sec:business-games}

\input{02_business_games/business_games.tex}

\input{02_business_games/gry_biznesowe.tex}

\subsection{Behavioral Approach to Management}
\label{sec:behavioral-approach}

\input{03_psychometry/psychometry.tex}

\section{The Project Game}
\label{sec:project-game}

\input{04_project_game/project_game.tex}

\section{Conclusion and Future Work}
\label{sec:conclusion}

\input{05_conclusion/conclusion.tex}


\bibliographystyle{splncs03}
\bibliography{%
01_introduction/citations,%
02_business_games/citations,%
03_psychometry/citations,%
04_project_game/citations,%
05_conclusion/citations%
}

\end{document}

%% file: 01_introduction/introduction.tex
%Here will be the contents of the introduction section
%As for now it may serve as a testing environment

%Bibliography remainder
%https://scholar.google.pl/
%Biblioteka -> Eksport do bibtexa
%Dodanie wpisów w pliku citations.bib w swoim folderze

%Some initial work on the subject of simulation \cite{okulewicz2016finding}.

Psychometric tests and business games have been both available for now over a several years as people and team performance assessment tools. Their existence allowed project managers, consultants and the teams themselves with identifying their working style, utilizing their strengths and avoiding their weaknesses in a day-to-day work.
Simulation information technology tools have been applied to various aspects of project management like team management \cite{okulewicz2016finding}, project plan construction \cite{waldzik2015risk} and building strategies for responses to security threats \cite{DBLP:conf/ecai/KarwowskiM16,DBLP:conf/icaisc/KarwowskiM15}.
Also, a general purpose multi-agent frameworks have been developed in order to design systems for modelling human behaviour \cite{singh2016integrating}.
We believe, that business games and psychometric analysis can be used in conjunction with the notion of a believable agents \cite{reilly1996}, to provide a more thorough overview of the project teams, and their effectiveness in completing various types of projects.
At the same time, a system combining the knowledge about the personality traits and action preferences in a business game, might be used for providing quantifiable data for socially inspired swarm algorithms and for testing level of differences between various types of personalities in the work-like environment.

The article starts with Section \ref{sec:business-games} presenting the introduction and applications of the business games. Section \ref{sec:behavioral-approach} provides an insight into the behavioural approach in management, with a special emphasis on a Belbin's Team Roles Inventory.
Section \ref{sec:project-game} proposes the Project Game, and describes its features, together with initial tests for a chosen Belbin's roles inspired software agents.
Finally, section \ref{sec:conclusion} concludes the paper.

%% file: 02_business_games/business_games.tex
According to Festus Oderanti \cite{dynamics2010}, business games are games that are put in the context of business strategies and are based on a game theory. Each player, or a team of players, faces a choice on every stage of the game, among two or more strategies. ''In business games, the conflicting interest of a firm may be to minimize the cost function, maximize the market share, or maximize the profit''. In the following section the term of business game is interchangeably used with a simulation or management game.

Business games broke into the market in the late 1950s \cite{keys1990role} and since 1970 they have become very popular among large companies and enterprise \cite{faria1998business}. Management game is often seen as a vehicle, which is used as a support to visualize and test strategy in a holistic approach. Moreover, it makes it possible to sketch the organizational cause-and-effect relationships and to communicate more clearly with structures that translate decisions into actions
%(Morecroft, 1999). Czy chodziło o Visualising and rehearsing strategy?
\cite{morecroft1999visualising}.
The five out of six studies have demonstrated that games are likely to teach their participants effectively, particularly when specific objectives are targeted \cite{randel1992effectiveness}.

In the past, board games were used only for entertainment purposes. Conversely, in 1970s a concept of serious game was invented by Clark Abt \cite{abt1987serious}, which made a new sense to the games. It was based on educational and psychological aspect of the board games. The current business games' market is rapidly developing. Simulation games are used for training, social and recruitment purposes. Advantages, among which are attractiveness of its form and the chance to get feedback on your own skills, make it highly valuable experience \cite{kim2017exploring}. Nevertheless, those games focus mostly on the final benefits and outcomes of the simulated world.

Many of the research studies have been applied to team performance in games, often with the assumption that high performing teams learn the most from a game experience \cite{keys1990role}. Recently, the veracity of this statement has been a subject of discussion \cite{henriksen2016can}. Beside this, the modern market offers an analysis of a personality, counting probability of adaptation employee to the specific environment of organization, identification of certain personal qualities, the visibility and ease of perception of the game, as well as appearing the hundreds of articles, assumptions and theories about the best combinations of roles in the group, but there is no product on the market that would put it all together.

The undeniable fact is that, team work tends to be more significant recently than ever before \cite{hbr2009teams}. The interactions between the team members are said to be the crucial key in cooperation and are likely to affect the efficiency of the projects the most \cite{google}. Furthermore, competences and predispositions of employee are need to be known to assign a role that suit him or her the best. Simulation games are widely known to be one of the methods of gathering information about competences of employees, what gives the opportunity to create the perfect team.

%% file: 02_business_games/gry_biznesowe.tex
%do scalenia i przetłumaczenia

%% file: 03_psychometry/psychometry.tex
Over the past few decades human behaviour and character traits has been a field of many studies. During those years different theories of personality have been developed, that a worth mentioning. One of the most popular has been proposed by Carl Jung \cite{jung1924psychological}. This theory distinguished two types of people’s attitude: extraversion and introversion. Other model of personality has been proposed by Hans Eysenck \cite{eysenck1950dimensions}. It is based on the biological perspective of personality and temperament theory. It includes 3 dimensions: extraversion, neuroticism and psychoticism. More relevant to this paper is one of the studies \cite{merrill1981personal}, which focused on human behaviour at work. The four proposed patterns of manner are: relational, technical, command and social specialists.
%https://www.cliffsnotes.com/study-guides/principles-of-management/the-evolution-of-management-thought/behavioral-management-theory
The works in that area of personality lead to development of a various psychometric tests \cite{carroll1974psychometric} , among which one of the currently popular is based on the Belbin's Team Roles Inventory \cite{Belbin}, which will be discussed in the next section.

\subsubsection{Belbin Team Roles Inventory}
\input{03_psychometry/belbin.tex}

\subsubsection{Modelling social behaviour with software agents}
Final area of interest, for constructing a business game in which people and software agents can interact, is the ability to model social behaviour.
Within that area it is important to notice the notion of a believable agent \cite{reilly1996}, and related research in swarm intelligence \cite{eberhart1995new,reynolds1987flocks}.
An interesting idea for modelling human behaviour within work teams is presented in \cite{martinez2010human}. The paper presents Team Knowledge-based Structuring (TEAKS) model of a virtual team. The team of software agents within TEAKS is configured using the characteristics of the real people. The model  produces statistical information obtained from the interaction between all the team members and their approximated performance with assigned tasks. Authors implemented in the model the most important in their opinion individual attributes that have an influence on human behaviour, which are:  creativity, emotions, personality traits and trust.
%This helped to allow the study and understanding of work team dynamics. 
Finally, it is important to mention the technical advancements in the area, coming with proposals of the technical frameworks supporting modelling such a behaviour \cite{singh2016integrating}.

%% file: 03_psychometry/belbin.tex
There has been many approaches to answer the question why some teams tend to end up with better results (including: time, costs, quality). One of the possible reasons is a proper composition of a roles of a team members, as proposes by a Belbin's theory of the \textit{Team Roles Inventory} \cite{belbin1981management,belbin2012team}. This theory defines Team Role as: ''A tendency to behave, contribute and interrelate with others in a particular way'' \cite{meredith2011management}. Belbin's research revealed, that it is important to pay more attention to the behaviour of teammates, than their solely their skills and intellect.
His research indicates, that the management is not only about selecting the right people for the position, but also matching people personalities with their job demands.
	The groups identified in the Team Roles Inventory, can be seen as further categorization of action-oriented, people-oriented and thinking-oriented people.
%It is valuable to state that one can be required to have different roles in different projects. But in most cases there is a preferred Team Role (often also the strongest).
	Every Team Role has some features, which under specific circumstances can be consider as a weakness. For example, thinking-oriented worker could be perceived as being unenthusiastic. In principle, Belbin's theory states that some foibles are the price that necessarily has to be paid for strength and calls them \textit{allowable weaknesses}.
 As a result he distinguishes nine Team Roles \cite{Belbin}.
\begin{description}
\item[Plant] – innovators, inventors, creative, independent, tend to be introverts that prefer to work alone, needed in initial stages of projects to provide different solutions for complex problems, counter-productive when too many in team.
\item[Resource Investigator] – enthusiastic as long as stimulated by others, extroverts, natural communicators, negotiators, easily develop new contacts, often good at promoting other people's ideas, usually see possibilities in new ideas, good at exploring and
\item[Coordinator] – mature, trusting, confident, spots individual talents, tend to set work towards shared goals, broad perspective, better in teams with diverse skills and equal rank.
\item[Shaper] – goal-oriented, determine to overcome obstacles, love to win and tend to be directive and decisive, often are considered as insensitive, even aggressive as they tend to push themselves and others, ideal managers to work under pressure and create action-oriented
\item[Monitor Evaluator] – serious-minded, individuals and do not like to be over-enthusiastic, slow in making decisions, often are considered as overthinking,  highly critical, tend to take all factors into account, unlikely to make intuitive decisions, good at analyzing problems and evaluating ideas.
\item[Implementer] – practical approach, high level of self-control and discipline, hard working, loyal, tend to do things as always, do what has to be done even if they do not like some tasks.
\item[Teamworker] – mild and sociable disposition, supportive, flexible, tend to adapt to different situations and people, good listeners, create harmony and avoid conflicts, but indecisive. Important because of their ability to solve interpersonal problems.
\item[Completer Finisher] – high attention to details, work with high concentration and high degree of accuracy, do not require external stimulus, can be trusted, finish tasks on time, create microculture where only acceptable standard is perfection.
9) Specialist – love to learn, accumulation of knowledge is their main motivation, like to be considered as experts that are worth asking for help and guidance, not natural team players but provide depth research.
\end{description}

%% file: 04_project_game/project_game.tex
%In the Project Game, the game is undertaken to gauge players attitude and strategy separately. The results are claimed to indicate both innate and acquired abilities of players. Due to the fact of considerable number of interactions between players and randomness of occurred events, it is strongly believed that natural behaviour and tendencies of players could be observed during the game, what leads to a better understanding of their roles in teams.
The main idea of the proposed Project Game \cite{theprojectgame2017} is to create a simulated work-like environment in which human and artificial agents could easily express various preferences in the areas of: (a) delivering quality, (b) delivering on-time, (c) level of risk acceptance, and (d) willingness to cooperate.
In order to evaluate teams and enabling team members expressing the above mentioned differences, the simulated projects are be ably to vary in their: (a) cost, (b) level of difficulty, (c) level of risks, and (d) competition pressure.

The proposed game model the project environment as a rectangular board, with appearing resources, on which the players can take a following set of actions with varying costs:
\begin{itemize}
	\item gathering resources
	\item testing products
	\item delivering products
	\item exchanging information
\end{itemize}
The game assumes a model of a self-organizing team, where the role of a leader is limited to the ability of executing information exchange.

\begin{figure}[t]
\includegraphics[height=0.33\textheight]{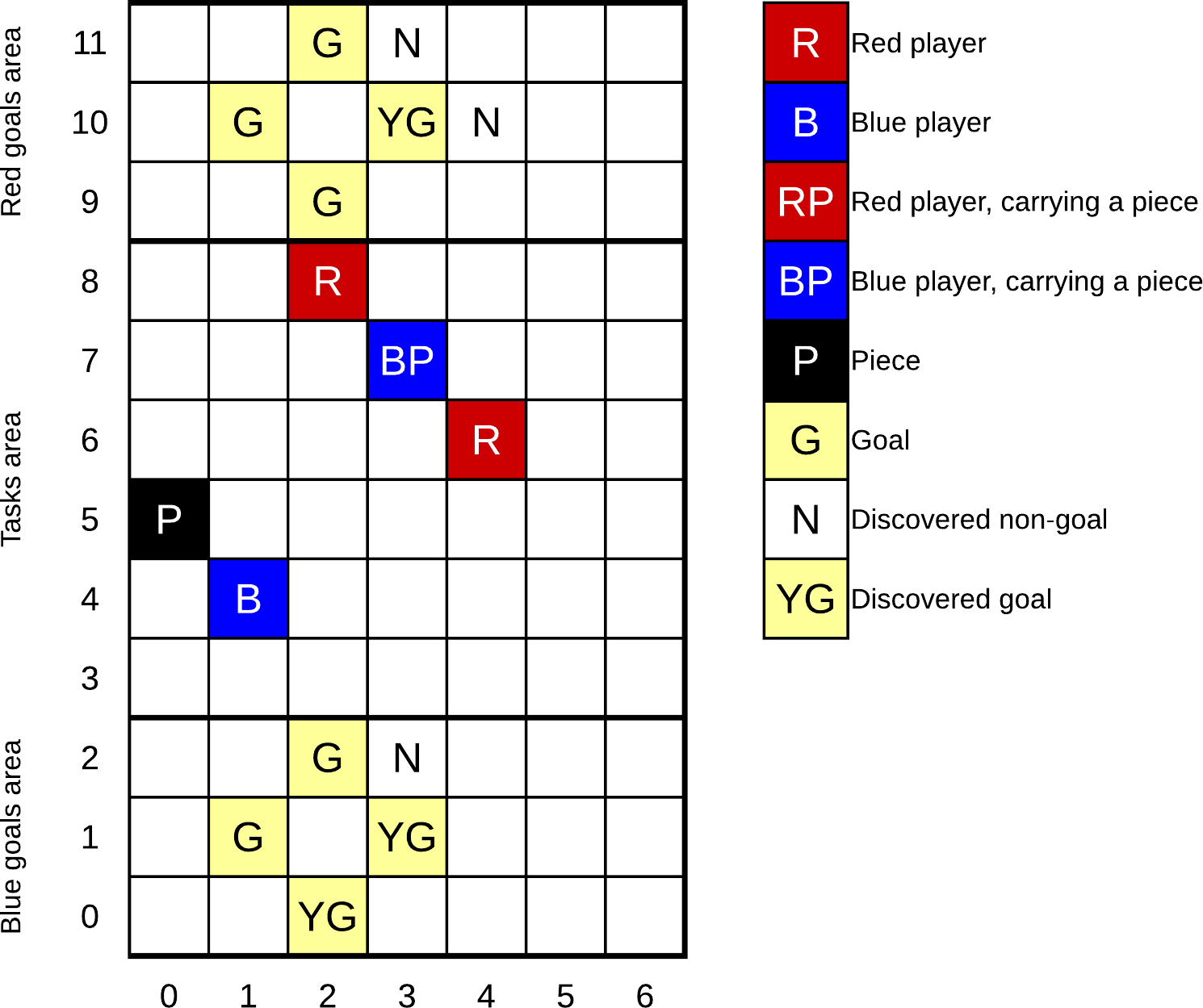}
\hfill
\includegraphics[height=0.33\textheight]{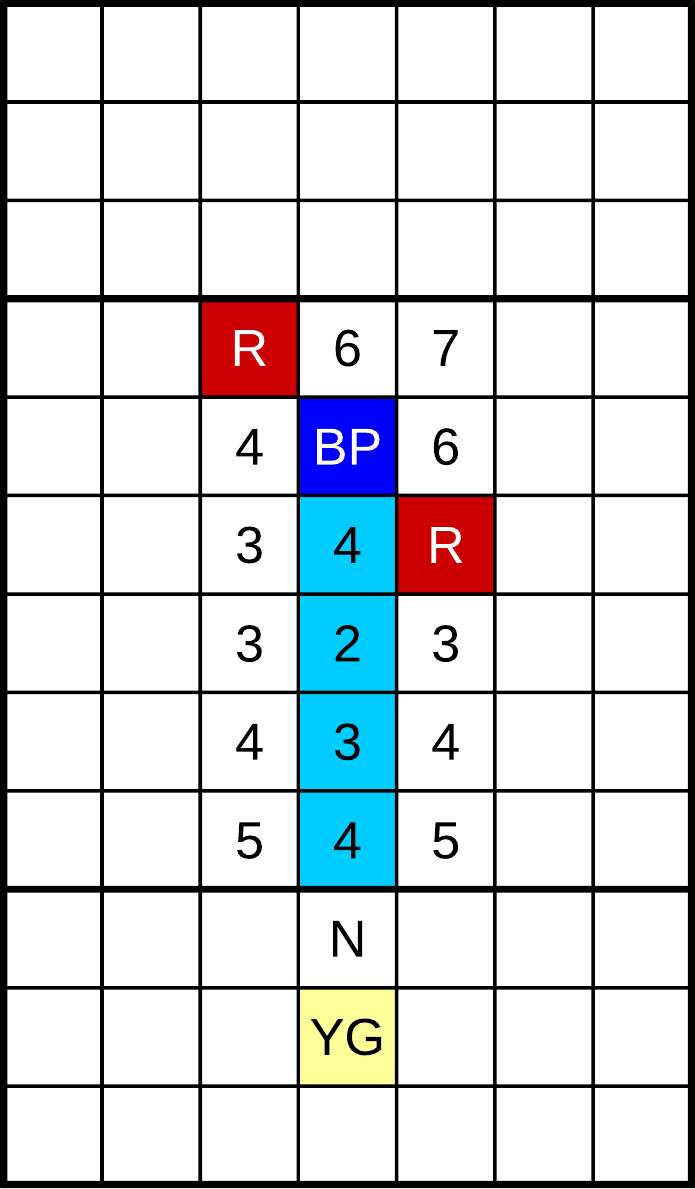}
\caption{Complete state of the game board (on the left) and its view by a single player (on the right).
\label{fig:gamestate}}
\end{figure}

The game is played by two teams of players (\texttt{red} and \texttt{blue}) on a board with three distinctive areas: \texttt{tasks area}, \texttt{red team goals area}, \texttt{blue team goals area} (see Figure \ref{fig:gamestate}).
The tasks area is where the resources randomly appear and from where might be gathered.
Goals area is where the products might be placed to validate if they have been developed according to the needs.
Figure \ref{fig:gamestate} presents a snapshot of a game state with a players view of it.
The game is won by the team which is first able to positively validate all the goals,
by delivering products for which no risk has materialized.

\begin{figure}[t]
\centering%
\includegraphics[width=0.95\textwidth]{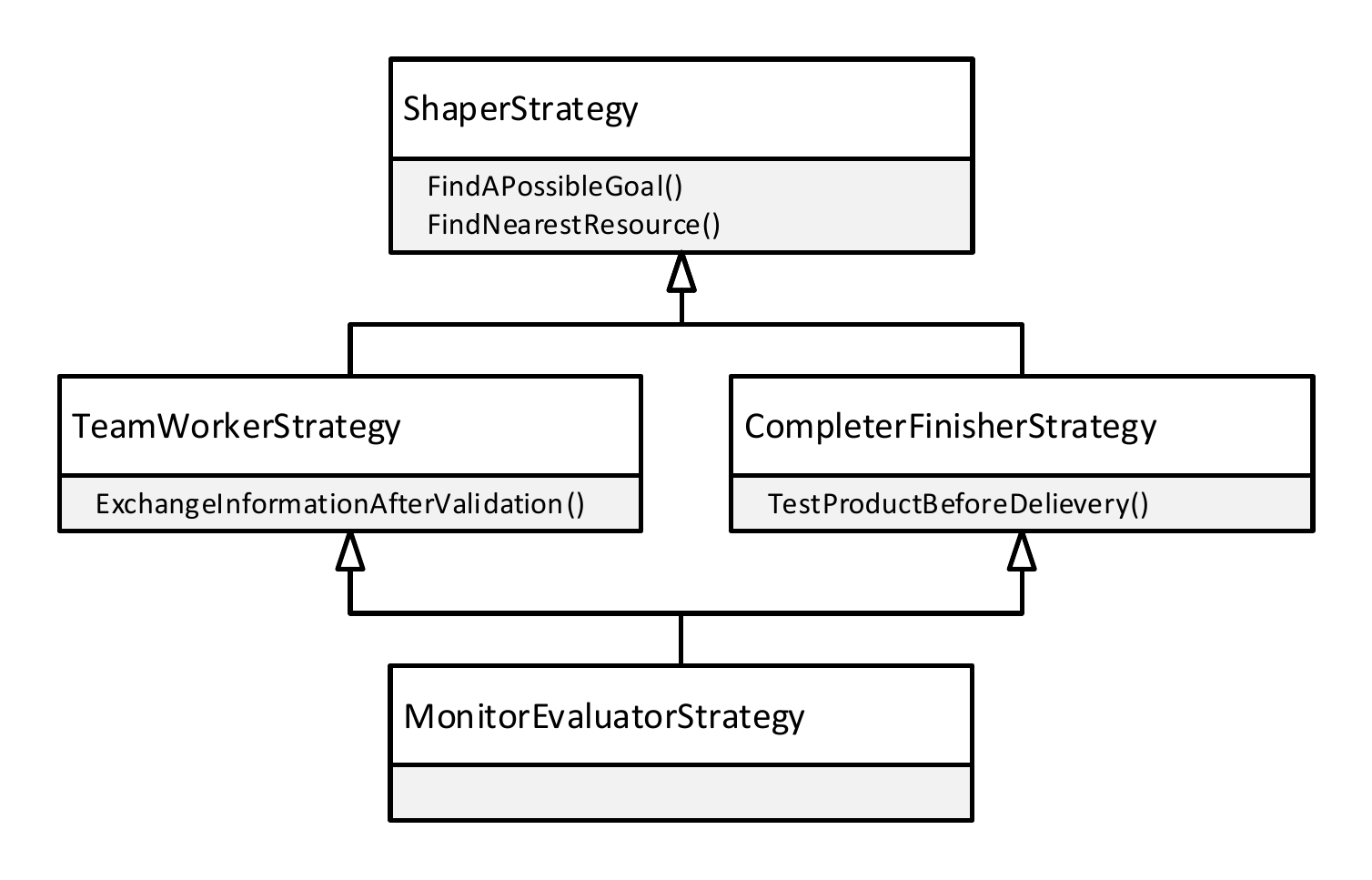}
\vspace{-1.5em}
\caption{Hierarchy of implemented strategies with a summary of their possible actions.
\label{fig:strategies}}
\end{figure}

\subsection{Players' Strategies}
In order to test the ability of the game to capture different effectiveness as a function of team member action preferences and project properties, four players' strategies have been proposed, loosely based on a selected Belbin's Team Roles.
The roles chosen for the tests included: \textit{Shaper}, \textit{Teamworker}, \textit{Completer Finisher} and \textit{Monitor Evaluator}, with Figure \ref{fig:strategies} giving a hierarchy of those Team Roles' strategies.

\subsubsection{Basic \textit{Shaper} Strategy}
main feature is maintaining the focus on completing the project as fast as possible under the optimistic assumptions.
The strategy chooses only the actions directly connected with project objective fulfilment: gathering resources and delivering products.
This strategy ignores the risks of delivering untested products and multiple discoveries of the same goal field by the different players.

\subsubsection{\textit{Completer Finisher} Strategy}
enhances the basic strategy with the ability to test the products before their delivery.

\subsubsection{\textit{Teamworker} Strategy}
enhances the basic strategy with the ability to exchange information with a randomly chosen team member, after a successful goal field state discovery.

\subsubsection{\textit{Monitor Evaluator} Strategy} simply combines the \textit{Completer Finisher} and \textit{Teamworker} behaviour. Player following that strategy is testing the products before their delivery
and sends requests to exchange information after goal validation. 

\subsection{Experiments}
In order to test the ability of the game to capture differences between the strategies
a set of matches between the 3 player teams of homogeneous agents has been played
within The Project Game environment \cite{theprojectgame2017}.
All the games have been played on a 6 by 18 fields boards with 4 goals for each of the teams.
The games varied with a level of risk associated with each of the resources.\footnote{The risk materialized either upon product testing or its validation.}

\begin{table}[t]
\caption{Action costs (counted in milliseconds) and values of the game parameters
\label{tab:params}}
\begin{center}%
\begin{tabular}{l|r}
\textbf{Action} & \textbf{Cost} \\\hline
Single field movement & 20 \\
Testing (initial) & 200 \\
Testing (final) & 50 \\
Information exchange & 300 \\
Discovery & 100 \\
Pick up & 20 \\
Place & 20 \\
\end{tabular}%
\hspace*{2em}
\begin{tabular}{l|r}
\textbf{Parameter} & \textbf{Value} \\\hline
Players & 3 \\
Risk & $\lbrace 0.0, 0.2, 0.5, 0.8 \rbrace$ \\
Board width & 6 \\
Board height & 18 \\
Task area height & 6 \\
Goals & 4 \\
Initial pieces & 6\\
Pieces appearing frequency & 300ms
\end{tabular}%
\end{center}
\end{table}

For each pair of strategies a several games have been played, with players of the same strategy taking different opposing sides. The detailed parameters of the games are presented in Table \ref{tab:params}.

\begin{figure}[t]
\centering%
\includegraphics[width=0.95\textwidth]{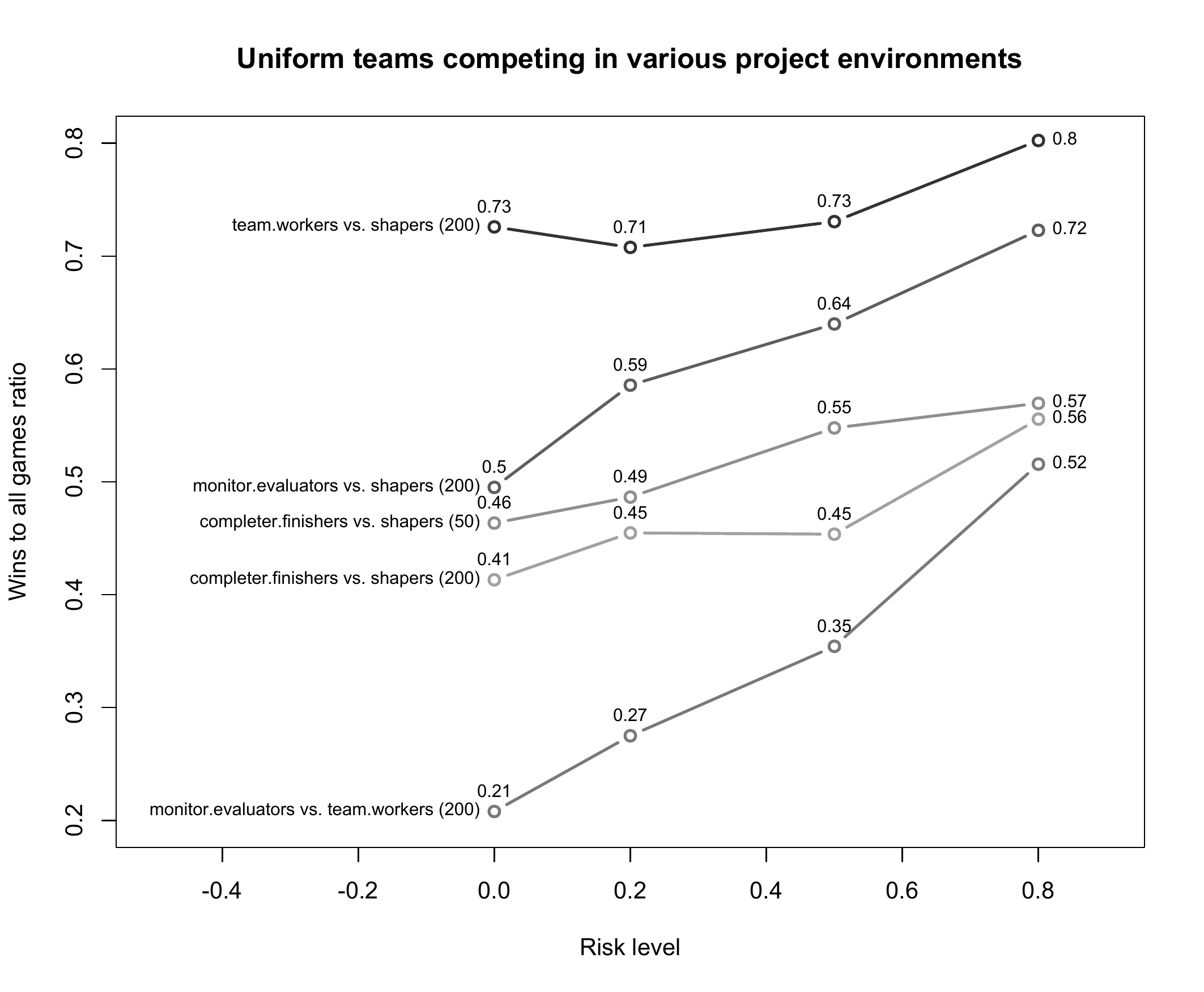}
\vspace{-1.5em}
\caption{Percentage of won games in various level of risk project environment for a games between a more complicated and a simpler strategies pairs. Testing cost in a given scenario is presented in the parentheses beside the strategies pairs names.
\label{fig:gameresults}}
\end{figure}

The percentage of the games won by a more complex strategy for selected matches is presented in Fig.~\ref{fig:gameresults}. It can be clearly observed that the strategies employing communication (i.e. Teamworker and Monitor Evaluator) have an advantage over the strategies that did not use communication (i.e. Shaper and Completer Finisher).
While percentage of the games won by the players utilizing ability to test products (i.e. Completer Finisher and Monitor Evaluator) grew with the larger levels of risk. Only after lowering testing cost from 200ms to 50ms Completer Finishers were able to gain advantage over Shapers.

%TODO The proposed environment allowed for differentiate between the projet size and level of risks with three distinct topologies: al-to-all, random-to-random, leader-to-all

%TODO Next edition needs to have the ability to destroy piece (i.e. fire people)

%% file: 05_conclusion/conclusion.tex
Proposed Project Game allowed for achieving various levels of advantages between sample strategy in relation to the project properties: level of risk and product testing costs. The game has also been able to capture a crucial role of communication within a project development.

Future work should consist of creating a user interface allowing human agents to participate in the game, and capturing human agent characteristics in relation to their psychometric tests results. 
Additional work, concerning ability to include reasoning, charisma and leadership abilities might be necessary in order to capture more features of a work-like environments.